\documentclass[10pt,prl,twocolumn,showpacs]{revtex4}
\usepackage{graphicx,amsmath,amssymb}
\begin{document}

\def\nn{\nonumber}
\def\kc#1{\left(#1\right)}
\def\kd#1{\left[#1\right]}
\def\ke#1{\left\{#1\right\}}
\renewcommand{\Re}{\mathop{\mathrm{Re}}}
\renewcommand{\Im}{\mathop{\mathrm{Im}}}
\renewcommand{\b}[1]{\mathbf{#1}}
\renewcommand{\c}[1]{\mathcal{#1}}
\renewcommand{\u}{\uparrow}
\renewcommand{\d}{\downarrow}
\newcommand{\bsigma}{\boldsymbol{\sigma}}
\newcommand{\blambda}{\boldsymbol{\lambda}}
\newcommand{\Tr}{\mathop{\mathrm{Tr}}}
\newcommand{\sgn}{\mathop{\mathrm{sgn}}}
\newcommand{\sech}{\mathop{\mathrm{sech}}}
\newcommand{\diag}{\mathop{\mathrm{diag}}}
\newcommand{\Pf}{\mathop{\mathrm{Pf}}}
\newcommand{\half}{{\textstyle\frac{1}{2}}}
\newcommand{\sh}{{\textstyle{\frac{1}{2}}}}
\newcommand{\ish}{{\textstyle{\frac{i}{2}}}}
\newcommand{\thf}{{\textstyle{\frac{3}{2}}}}
\newcommand{\SUN}{SU(\mathcal{N})}
\newcommand{\N}{\mathcal{N}}

\title{String theory embeddings of non-relativistic field theories and their holographic Ho\v rava gravity duals}

\author{Stefan Janiszewski and Andreas Karch}

\affiliation{Department of Physics, University of Washington, Seattle, WA
98195-1560, USA}

\date\today

\begin{abstract}
We argue that generic non-relativistic quantum field theories have a holographic description in terms of Ho\v rava gravity. We construct explicit examples of this duality embedded in string theory by starting with relativistic dual pairs and taking a non-relativistic scaling limit.
\end{abstract}


\maketitle

{\bf Introduction:} Holography \cite{Maldacena:1997re,Gubser:1998bc,Witten:1998qj} asserts an equivalence between quantum gravity on asymptotically anti-de Sitter (AdS) spacetimes and relativistic quantum field theories (QFTs). Holography has profound consequences for both sides of the duality. Relativistic QFTs are defined non-perturbatively via a path integral, which, in principle, can be simulated on a computer when space and time are discretized. Via holography they provide a complete definition of quantum gravity. Holography can also be used to study otherwise uncontrollable dynamical properties of these field theories at strong coupling, where the dual gravitational theory is dominated by a classical saddle point.

In this work we will show that a similar relation connects non-relativistic (NR) QFTs, that is, many body quantum mechanical systems, to a recently proposed alternate theory of gravity, Ho\v rava gravity \cite{Horava:2009uw}.
The guiding principle that allows us to establish a dictionary between the two sides is symmetry. A large class of NR QFTs share a common set of symmetries: translations, rotations, Galilean boosts, and a conserved particle number. Important examples in this class include quantum Hall systems as well as the unitary Fermi gas. We show that these symmetries demand that a putative holographic dual of any such system
has to be of the Ho\v rava gravity type. Embedding this construction in string theory provides explicit dual pairs. In this way a NR QFT and its Ho\v rava gravity dual arise as non-relativistic limits of a known holographic duality. As the strongly correlated electron systems of interest in condensed matter physics are intrinsically non-relativistic, this NR holography should pave the way for applications of holographic techniques to more realistic systems.

Ho\v rava gravity, like Einstein gravity, is a metric theory built around invariance under diffeomorphisms, that is, coordinate changes. Unlike Einstein gravity, it insists on a preferred temporal slicing. The only changes of coordinates that are symmetries are ``foliation preserving diffeomorphisms" (FDiffs): (time dependent) changes of the spatial coordinates and space independent reparametrization of time. Due to the reduced symmetry, Ho\v rava gravity has an extra scalar mode. The less restrictive symmetry also allows extra terms in the action. The low energy action is not necessarily the standard Einstein-Hilbert (EH) term but has two additional free parameters \footnote{There exist truncations of Ho\v rava gravity where one of the metric factors is allowed to depend only on time, not space. In this ``projectable" version of Ho\v rava gravity the low energy theory only has one free parameter. In this work we will only consider the ``non-projectable" theory.}. This makes it difficult to reconcile Ho\v rava gravity with observed gravitational phenomena of our world. However, it has exactly the right properties to be the dual description of a generic NR QFT.

The reason Ho\v rava gravity has been intensely studied in recent years is that it has a much nicer ultraviolet (UV) behavior than Einstein gravity. For a rescaling of energy, spatial and temporal derivatives do not have to scale in the same way. Under such an anisotropic scaling the action of Ho\v rava gravity can be seen to be power counting renormalizable making the existence of UV fixed points possible. Some evidence for such fixed points has been seen in recent lattice studies based on dynamical triangulation \cite{Anderson:2011bj}. By giving an explicit embedding of Ho\v rava gravity into string theory we show that at least some of its incarnations are, in fact, consistent quantum theories of gravity. As Ho\v rava gravity is potentially already UV complete on its own, this new duality may help understand holography beyond the classical gravity limit (which typically is equivalent to a large number of colors $N_c$ in the QFT).

This work indicates that the holographic dualities known so far between relativistic QFTs and Einstein gravity are just the tip of the iceberg. In the large landscape of consistent quantum mechanical theories, relativistically invariant theories are only a very special, highly constrained corner. This holographic duality suggests that the well established relativistic quantum theories of gravity realized as string theories are similarly just special corners of a large landscape of Ho\v rava-like quantum gravities.

{\bf Global, local and spurionic symmetries:} Symmetries form a cornerstone of our understanding of modern physics. There are, however, some important distinctions to be made. The standard model of particle physics has a Lagrangian based on an $SU(3) \times SU(2) \times U(1)$ gauge symmetry. Despite their name, gauge symmetries aren't really symmetries; they are redundancies in our chosen description. Physical reality is only described by gauge invariant observables. To obtain a simple Lagrangian, extra redundant degrees of freedom are introduced together with an equivalence relation (the gauge symmetry) that allows us to set to zero all unphysical degrees of freedom. This well known point has become widely appreciated after Seiberg's discovery \cite{Seiberg:1994pq} of dualities in ${\cal N}=1$ supersymmetric gauge theories where different gauge groups can be shown to describe one and the same physical theory. In contrast, a global symmetry is a true symmetry of physical observables. Physical quantities furnish a representation of the global symmetry. Most importantly, global symmetries imply conservation laws.

Standard global symmetries act on the dynamical fields in a theory and leave the action invariant. It is however often useful to consider global transformations that are only symmetries if in addition to the dynamical fields some of the parameters in the Lagrangian transform. One example of this is a chiral rotation in the theory of massive Dirac fermions. For a massless fermion, the left handed and right handed components of the fermion can be rotated independently. A mass term in the Lagrangian spoils this symmetry unless one lets the mass parameter transform under the symmetry as well. Such ``spurionic" symmetries do not give rise to conservation laws, but they still put powerful constraints on physics. Most importantly, spurionic symmetries constrain the dependence of the low energy effective Lagrangian ($L_{eff}$) on microscopic parameters.

Diffeomorphisms in a relativistic QFT are an example of such a global spurionic symmetry (despite the fact that they can be performed locally at every point in space and time). In a QFT we should think of the background spacetime metric as a set of coupling constants (5 at each point in space and time for $d=3$ spatial dimensions); it is not a dynamical field we solve for. Diffeomorphisms $\delta x^{\mu} = -\xi^{\mu}(x^{\nu})$ leave the theory invariant as long as they act on these background couplings in the usual way:
\begin{equation}
\delta g_{\mu \nu} = \xi^{\lambda} \partial_{\lambda} g_{\mu \nu} +
 g_{\lambda \nu} \partial_{\mu} \xi^{\lambda} + g_{\mu \lambda} \partial_{\nu} \xi^{\lambda} .
\end{equation}
True global symmetries arise as the subgroup of the spurionic symmetries that leave a particular background invariant. E.g. for the trivial background $g_{\mu \nu} = \eta_{\mu \nu}$ these are boosts, rotations and translations. Nevertheless, the full spurionic symmetries give powerful new constraints on $L_{eff}$. For example, consider a gapped system on a compact, curved background space. The system has a non-trivial Casimir energy depending on the geometry of the background. As long as the curvature radii of the background are large in units set by the gap, we can write down $L_{eff}$ by including all 2-derivative terms consistent with symmetries, and derive the Casimir energy. As all dynamical degrees of freedom are gapped, the only field appearing in $L_{eff}$ is the background metric itself. The spurionic symmetry determines that the metric can only appear in $L_{eff}$ in diffeomorphism invariant combinations: a cosmological constant and the standard EH term.

Son and Wingate \cite{Son:2005rv} have demonstrated that a large class of NR QFTs have time dependent spatial diffeomorphisms, together with local $U(1)$ transformations, as global spurionic symmetries acting on the background metric and the background electric and magnetic fields encoded in the vector potential $A_{\mu}(t,\vec{x})$. These symmetries can easily be seen for free fields, but they are preserved in many phenomenologically important interacting systems, in particular the fractional quantum Hall effect and the unitary Fermi gas. In \cite{Son:2005rv} the full FDiff group (which includes the additional time reparametrization) is only a symmetry if the system is conformally invariant, but it can be realized as a spurionic symmetry in any NR QFT with spatial diffeomorphism invariance by introducing \cite{Son:2008ye} one more scalar background field $e^{-\Phi}$ which enters the Lagrangian as a source for energy density \footnote{The full relativistic diffeomorphism invariance can in fact be realized as a spurionic symmetry by introducing yet another additional vector $\vec{B}$ coupled to the energy current. Here we will restrict ourselves to the case $\vec{B}=0$.}. Under spatial diffeomorphisms $\delta x^i = -\xi^i(t,\vec{x})$, time reparametrizations $\delta t = -f(t)$, and $U(1)$ rotations $\lambda(t,\vec{x})$ the backgrounds transform as:
\begin{eqnarray}
\delta A_t &=& - \dot{\lambda} + \dot{f} A_t  + \dot{\xi}^i A_i + f \dot{A}_t+ \xi^{j} \partial_{j} A_t  \nonumber \\
\delta A_i &=& -\partial_i \lambda + m e^{\Phi}g_{ij} \dot{\xi}^j + f \dot{A}_i + \xi^{j} \partial_{j} A_i + A_j \partial_i \xi^j
\nonumber \\
\delta \Phi &=& -\dot{f} + f \dot{\Phi} +\xi^{j} \partial_{j} \Phi \nonumber \\
\delta g_{ij} &=& g_{ik} \partial_j \xi^k  + g_{kj} \partial_i \xi^k
+ f \dot{g}_{ij} + \xi^{k} \partial_{k} g_{ij}.
\label{trafos}
\end{eqnarray}

These symmetries strongly constrain $L_{eff}$. In quantum Hall systems they fix both the Hall viscosity as well as the current response to a spatially varying electric field \cite{Hoyos:2011ez}. A simple way to derive these transformation rules is to start with the theory of a mass $m$ free relativistic scalar in the presence of a chemical potential $\mu=m c^2$. Dropping terms with inverse powers of $c$ from the action reduces the free relativistic system to a free NR one. Spatial diffeomorphisms remain a symmetry while subleading temporal diffeomorphisms reduce to the local spurionic $U(1)$ symmetry associated with particle number.

The trivial background $g_{ij}=\delta_{ij}$, $A_{\mu}=\phi=0$ is left invariant by translations ($\xi^i=const.$), rotations ($\xi^i = M_{ij} x^j$ with $M_{ij}=-M_{ji}$), and Galilean boosts ($\xi^i=v^i t$, $\lambda=\vec{v} \cdot \vec{x}$). These are hence genuine global symmetries of our system. Note that allowing time dependent spatial diffeomorphisms is important in order to realize Galilean boosts. If, in addition, the QFT has local Weyl rescalings
\begin{equation}
\label{weyl} \delta_\omega g_{ij} = 2 \omega g_{ij}, \quad \quad \delta_\omega \Phi = -2 \omega
\end{equation}
as a spurionic global symmetry (as is the case e.g. in the unitary Fermi gas), two additional global symmetries arise for the trivial background:
scale ($\omega=-\kappa$, $f=2 \kappa t$, $\xi^i=\kappa x^i$), and special conformal ($\omega= -C t$, $f = C t^2$, $\xi^i = C t x^i$, $\lambda= C \vec{x}^2/2$) transformations.

{\bf Holographic dictionary:} One of the key pieces of evidence for the holographic equivalence is the matching of global symmetries. This has to include the spurionic global symmetries. In a gravitational theory diffeomorphisms vanishing at the boundary of AdS are a gauge symmetry, hence a redundancy, and not a physical symmetry. The spurionic global symmetries correspond to large diffeomorphisms that do not vanish at the boundary. In normal coordinates, where the bulk metric takes the form $ds^2 =r^{-2}(dr^2 + g_{\mu \nu} dx^{\mu} dx^{\nu})$, with $g_{\mu \nu}$ finite at $r=0$, these global spurionic symmetries correspond to diffeomorphisms $\xi^{\mu}$ independent of the holographic coordinate $r$: $\xi^r$ is gauged fixed by going to normal coordinates; $\xi^{\mu}$ with positive powers of $r$ are gauge symmetries and hence redundancies; negative powers of $r$ would destroy the asymptotic AdS form.

Following this logic, a holographic description of a generic NR QFT requires a gravitational theory built around FDiffs instead of relativistic diffeomorphisms; $r$ independent FDiffs in the bulk can account for the spurionic global symmetry of the boundary QFT. Therefore, symmetries suggest that the dual to the generic NR QFT is Ho\v rava gravity on asymptotically AdS space. Writing the bulk spacetime metric in terms of the lapse $N$, shift $N_I$ ($I$ runs over the $d$ spatial indices $i$ as well as the radius $r$), and spatial metric $G_{IJ}$, the extra scalar mode of Ho\v rava gravity is $N_r$; the bulk metric can no longer be brought into normal form. As mentioned, the action of Ho\v rava gravity can be taken to be that of Einstein gravity with extra terms. At the two derivative level we can add $\tilde{\lambda} K^2$ (where $K$ is the trace of the extrinsic curvature of the spatial slice) and $\alpha (\nabla_I N)^2/N^2$ to the Lagrangian. The free couplings $\tilde{\lambda}$ and $\alpha$ parametrize the deviations from the EH action. If $\alpha$ and $\tilde{\lambda}$ vanish the kinetic term for the extra scalar mode vanishes identically and the limit is degenerate. They can however be parametrically small and still lead to a healthy version of Ho\v rava gravity \cite{Blas:2009qj}. In our explicit string theory example these two parameters vanish at tree level and will only be generated at loop level (at order 1 in the large $N_c$ counting whereas the coefficient of the EH term is of order $N_c^2$). We will refer to the limit of parametrically small $\alpha$ and $\tilde{\lambda}$ as the ``probe limit".

The connection between Ho\v rava and Einstein gravity is easiest to see in the khronon formalism \cite{Blas:2009yd,Germani:2009yt}. Starting from Ho\v rava gravity one introduces an auxiliary scalar field, the khronon, and assigns it transformation properties under temporal diffeomorphisms that compensate the non-invariance of Ho\v rava gravity. One can always go back to Ho\v rava gravity by fixing ``unitary gauge" in which the extra khronon field is set to $c^2 t$ and its fluctuations are absent. Temporal diffeomorphisms are explicitly broken. A completely equivalent way to think about Ho\v rava gravity with the khronon is to treat it as Einstein gravity coupled to a scalar field. In this language the $c^2 t$ background profile for the scalar picks a preferred time slicing, while $\alpha$ and $\tilde{\lambda}$ appear as parameters in the scalar action. The khronon formalism is especially convenient in the probe limit: as the prefactor of the khronon action is parametrically small, the energy density of the khronon doesn't backreact on the metric. The khronon profile simply imprints a preferred notion of time on a fixed background spacetime. In the probe limit, any solution of Einstein gravity descends to a solution of Ho\v rava gravity.

The scalar khronon also naturally gives rise to the extra spurionic $U(1)$ symmetry of the putative dual QFT. As first introduced in \cite{Horava:2009vy}, the $U(1)$ is realized as a subleading (in $c$) term in the temporal diffeomorphisms. However, the $U(1)$ acts as shifts on the khronon fluctuations. This is simply a reflection of the fact that the extra scalar mode can be shuffled back and forth between the khronon and $N_r$. By choosing unitary gauge (that is by setting the fluctuating khronon to zero) we completely fix the $U(1)$ invariance in the bulk. No global spurionic $U(1)$ transformation acting on the boundary emerges. This has to be contrasted with the case of a bulk $U(1)$ gauge invariance in relativistic holography, where the gauge choice $A_r=0$ for the corresponding gauge field leaves $r$-independent gauge transformations as a residual symmetry, reproducing the corresponding spurionic symmetry of the dual QFT. The scalar khronon theory has additional problems. To recover the full FDiff invariance, the scalar has to have a field reparametrization invariance. No exact global symmetries exist in quantum gravity and so the shift and reparametrization invariance we have to demand from the khronon field could at best exist at the classical level. In addition, a scalar field with a background growing linear in time corresponds to a uniform energy density, such a background is likely subject to clumping instabilities.

All these problems can be solved by working with a massless vector khronon $A_M$. This can be achieved by starting with Einstein-Maxwell gravity and introducing a background with a constant $A_t=c^2$. This choice yields a preferred time slicing. Taking the large $c$ limit and dropping all terms with negative powers of $c$ from the transformations of the Einstein-Maxwell fields reduces the content to that of Ho\v rava gravity. As $A_t=c^2$ is pure gauge, it does not correspond to any energy density and so is not subject to clumping. Since the corresponding symmetry is gauged it is compatible with quantum gravity. The $U(1)$ spurionic symmetry of the NR QFT is naturally realized as the gauge invariance acting on this vector khronon. It is straightforward to verify \cite{long} that in the large $c$ limit certain combinations of the bulk fields
transform exactly as the boundary sources of the dual QFT in eq.~(\ref{trafos}).

By starting with classical Einstein-Maxwell theory and turning on $A_t=c^2$ we naturally end up with Ho\v rava gravity in the probe limit. $\alpha$ and $\tilde{\lambda}$ are allowed by the symmetries and will be generated by loops, as will a spatial kinetic term $\beta F_{IJ} F^{IJ}$ for the gauge field. In this probe limit, any solution of Einstein-Maxwell gravity naturally descends. In particular, as long as we have a negative cosmological constant we can study this theory on pure AdS or an AdS black hole.

This bulk theory realizes the spurionic symmetry eq.~(\ref{trafos}) of a NR QFT via $r$-independent transformations acting on the boundary sources. Spurionic Weyl rescalings from eq.~(\ref{weyl}), get implemented as radial diffeomorphisms ($\xi^r= -\omega r$, $\xi^{i} = r^2 g^{ij} \partial_j \omega/2$). Scale transformations correspond to constant $\omega$. For a pure AdS bulk, $N_I=0$, $N=r^{-1}$, $G_{IJ}=r^{-2} \delta_{IJ}$ the on-shell bulk action is only rescaled by an overall constant. The dual field theory is scale invariant: while the gravity fields transform non-trivially, the combinations of bulk fields corresponding to the field theory quantities \cite{long} have scaling as an isometry. The field theory quantities have an additional isometry involving $\omega=-Ct$, corresponding to the special conformal transformation. The discrepancy between generic bulk fields and the special combinations that map to field theory sources can be understood by examining the action of subleading temporal diffeomorphisms on the fields. The field theory quantities are invariant under this transformation, while the bulk gravity fields generically transform. This leads to two options. Generally, one can consider Ho\v rava gravity on AdS with its FDiff symmetry as dual to a NR QFT with scale invariance. The additional bulk fields map to field theory operators that respect scaling but do not respect the conformal symmetry. As long as FDiff invariant bulk actions are used, correlators and other quantities calculated via the correspondence will exhibit scale symmetry. Alternatively, in order to incorporate the special conformal invariance of Schr\"odinger symmetry the symmetries of the bulk must be expanded beyond FDiffs to give an action invariant under the subleading temporal diffeomorphisms. In this case the additional bulk fields are rendered redundant and only the conformally invariant field theory quantities are physical. An example of a bulk theory with this extra gauge symmetry is the covariant Ho\v rava gravity of \cite{Horava:2009vy} coupled to a Maxwell field.

The scale symmetry comes with dynamical critical exponent $z=2$, which is required to keep the source $\Phi=0$. It generically also comes with a non-vanishing hyperscaling violating critical exponent \cite{Huijse:2011ef} $\theta=1$. For backgrounds with vanishing extrinsic curvature, the bulk Lagrangian under scaling only picks up an overall factor of $\omega$ from the prefactor of $N$ in the measure. This gives rise to $\theta=1$ scaling of the free energy density. This non-invariance of the Lagrangian can be compensated if the whole Lagrangian is proportional to a scalar field $e^{\sigma}$ which itself shifts by $\delta \sigma=-\omega$. We will see this to be the case in our first explicit string theory model. In this special case one has $\theta=0$.

These backgrounds dual to scale invariant NR QFTs allow for a highly non-trivial check of our proposal: the correlation function of a charge $q$, dimension $\Delta$ scalar operator in a NR QFT is strongly constrained \cite{Henkel:1993sg} by the combination of scale invariance and Galilean boost invariance \footnote{Note that this is true even at non-zero $\theta$, as boosts and the relative scaling between $\vec{x}$ and $t$ fix the 2-pt functions. In contrast, the hyperscaling violating backgrounds of \cite{Ogawa:2011bz,Huijse:2011ef} are not boost invariant and hence exhibit substantially more complicated 2-pt functions.},
$\langle O(t,\vec{x}) O(0) \rangle \sim t^{-\Delta} e^{- q \frac{|\vec{x}|^2}{2t}}$. Putting a charge $q$ scalar field $X$ on an AdS background with vector khronon, we note that its leading kinetic term is no longer the 2-derivative $|\partial_0 X|^2$ but instead the 1-derivative $q \Im (X^* A_0 \partial^0 X)$. Dropping the 2-derivative term (which is a higher derivative correction) the action of the scalar $X$ is identical to that of a probe scalar on the Schr\"odinger geometry \cite{Balasubramanian:2008dm,Son:2008ye} and hence leads to a correlator consistent with symmetry. Note however that generically we also generate a superleading order $c^2|X^2|$ term in the action from the $A_t^2$ pieces in the covariant derivatives. This has to cancel against a bulk mass term for $X$, just as it did in the field theory. Otherwise the equations of motion set $X=0$ identically. Generically the NR limit will kill all excitations; one needs fields with a particular value of the bulk mass to have any non-trivial dynamics left over. Only a small subset of bulk fields survives the limit.

{\bf Embeddings in string theory:}
In order to embed Ho\v rava gravity coupled to a Maxwell field into a relativistic setting all we need to do is to study Einstein-Maxwell in the presence of a background gauge field $A_t=m c^2$ (where $m$ is the mass of at least one charged field in the dual field theory) and take the large $c$ limit. As we have seen, in the bulk we need charged fields to get a mass linked to $m$ as well in order to survive the limit. One way to accomplish this is to use the bulk scalar field $e^{\Sigma}$ dual to the mass operator in the field theory and couple it via $e^{2 \Sigma} |X|^2$ to the bulk fields. The simplest example where we can easily realize this is a circle compactification. Starting with ${\cal N}=4$ SYM in 3+1 dimensions, we can get a massive $d=2$ theory by compactifying $x_3$ on a circle of radius $R=(mc)^{-1}$. The $d=2$ theory has a tower of excitations of mass $q m$ with integer charge $q$ under the $U(1)$ global symmetry we inherit from translations along the circle direction. The dual background geometry is AdS$_5$ written as $ds^2=r^{-2}\left [ (\frac{dx_3}{mc} + A_{\mu} dx^{\mu})^2  +
g_{\mu \nu} dx^{\mu} dx^{\nu} \right ]$. Now setting $A_t=cA_0=c^2 m$ and taking the $c$ to infinity limit we end up with a compact light-like circle; latter is known to have the full symmetries of a NR CFT \cite{Son:2008ye,Balasubramanian:2008dm,Goldberger:2008vg} with $\theta=0$. The $\sqrt{g_{33}}$ factor in the action, from the compactified viewpoint, plays the role of $e^\sigma$ and prevents hyperscaling violation.
The NR limit of setting the chemical potential equal to the mass and sending $c$ to infinity is nothing but the well known Seiberg/Sen limit \cite{Seiberg:1997ad,Sen:1997we} that views light-like compactifications as a zero radius limit of space-like ones.

A different set of string theory embeddings that allow a NR limit along the lines we describe here is provided by probe flavor branes \cite{Karch:2002sh}. With flavor branes it is once more easy to introduce a mass term for the charged fields in the relativistic QFT, and so we can once more introduce $A_t =m c^2$ and take the large $c$ limit. This limit was studied in \cite{Karch:2007br} and recently in more detail in \cite{Ammon:2012je}, where it was shown to give rise to a scale invariant theory with $z=2$ and $\theta=1$, consistent with our discussion of scaling above.

\bibliography{NRhorava2}

\begin{thebibliography}{24}
\expandafter\ifx\csname natexlab\endcsname\relax\def\natexlab#1{#1}\fi
\expandafter\ifx\csname bibnamefont\endcsname\relax
  \def\bibnamefont#1{#1}\fi
\expandafter\ifx\csname bibfnamefont\endcsname\relax
  \def\bibfnamefont#1{#1}\fi
\expandafter\ifx\csname citenamefont\endcsname\relax
  \def\citenamefont#1{#1}\fi
\expandafter\ifx\csname url\endcsname\relax
  \def\url#1{\texttt{#1}}\fi
\expandafter\ifx\csname urlprefix\endcsname\relax\def\urlprefix{URL }\fi
\providecommand{\bibinfo}[2]{#2}
\providecommand{\eprint}[2][]{\url{#2}}

\bibitem[{\citenamefont{Maldacena}(1998)}]{Maldacena:1997re}
\bibinfo{author}{\bibfnamefont{J.~M.} \bibnamefont{Maldacena}},
  \bibinfo{journal}{Adv.Theor.Math.Phys.} \textbf{\bibinfo{volume}{2}},
  \bibinfo{pages}{231} (\bibinfo{year}{1998}), \eprint{hep-th/9711200}.

\bibitem[{\citenamefont{Gubser et~al.}(1998)\citenamefont{Gubser, Klebanov, and
  Polyakov}}]{Gubser:1998bc}
\bibinfo{author}{\bibfnamefont{S.}~\bibnamefont{Gubser}},
  \bibinfo{author}{\bibfnamefont{I.~R.} \bibnamefont{Klebanov}},
  \bibnamefont{and} \bibinfo{author}{\bibfnamefont{A.~M.}
  \bibnamefont{Polyakov}}, \bibinfo{journal}{Phys.Lett.}
  \textbf{\bibinfo{volume}{B428}}, \bibinfo{pages}{105} (\bibinfo{year}{1998}),
  \eprint{hep-th/9802109}.

\bibitem[{\citenamefont{Witten}(1998)}]{Witten:1998qj}
\bibinfo{author}{\bibfnamefont{E.}~\bibnamefont{Witten}},
  \bibinfo{journal}{Adv.Theor.Math.Phys.} \textbf{\bibinfo{volume}{2}},
  \bibinfo{pages}{253} (\bibinfo{year}{1998}), \eprint{hep-th/9802150}.

\bibitem[{\citenamefont{Horava}(2009)}]{Horava:2009uw}
\bibinfo{author}{\bibfnamefont{P.}~\bibnamefont{Horava}},
  \bibinfo{journal}{Phys.Rev.} \textbf{\bibinfo{volume}{D79}},
  \bibinfo{pages}{084008} (\bibinfo{year}{2009}), \eprint{0901.3775}.

\bibitem[{\citenamefont{Anderson et~al.}(2012)\citenamefont{Anderson, Carlip,
  Cooperman, Horava, Kommu et~al.}}]{Anderson:2011bj}
\bibinfo{author}{\bibfnamefont{C.}~\bibnamefont{Anderson}},
  \bibinfo{author}{\bibfnamefont{S.~J.} \bibnamefont{Carlip}},
  \bibinfo{author}{\bibfnamefont{J.~H.} \bibnamefont{Cooperman}},
  \bibinfo{author}{\bibfnamefont{P.}~\bibnamefont{Horava}},
  \bibinfo{author}{\bibfnamefont{R.~K.} \bibnamefont{Kommu}},
  \bibnamefont{et~al.}, \bibinfo{journal}{Phys.Rev.}
  \textbf{\bibinfo{volume}{D85}}, \bibinfo{pages}{044027}
  (\bibinfo{year}{2012}), \eprint{1111.6634}.

\bibitem[{\citenamefont{Seiberg}(1995)}]{Seiberg:1994pq}
\bibinfo{author}{\bibfnamefont{N.}~\bibnamefont{Seiberg}},
  \bibinfo{journal}{Nucl.Phys.} \textbf{\bibinfo{volume}{B435}},
  \bibinfo{pages}{129} (\bibinfo{year}{1995}), \eprint{hep-th/9411149}.

\bibitem[{\citenamefont{Son and Wingate}(2006)}]{Son:2005rv}
\bibinfo{author}{\bibfnamefont{D.}~\bibnamefont{Son}} \bibnamefont{and}
  \bibinfo{author}{\bibfnamefont{M.}~\bibnamefont{Wingate}},
  \bibinfo{journal}{Annals Phys.} \textbf{\bibinfo{volume}{321}},
  \bibinfo{pages}{197} (\bibinfo{year}{2006}), \eprint{cond-mat/0509786}.

\bibitem[{\citenamefont{Son}(2008)}]{Son:2008ye}
\bibinfo{author}{\bibfnamefont{D.}~\bibnamefont{Son}},
  \bibinfo{journal}{Phys.Rev.} \textbf{\bibinfo{volume}{D78}},
  \bibinfo{pages}{046003} (\bibinfo{year}{2008}), \eprint{0804.3972}.

\bibitem[{\citenamefont{Hoyos and Son}(2011)}]{Hoyos:2011ez}
\bibinfo{author}{\bibfnamefont{C.}~\bibnamefont{Hoyos}} \bibnamefont{and}
  \bibinfo{author}{\bibfnamefont{D.~T.} \bibnamefont{Son}}
  (\bibinfo{year}{2011}), \bibinfo{note}{* Temporary entry *},
  \eprint{1109.2651}.

\bibitem[{\citenamefont{Blas et~al.}(2010)\citenamefont{Blas, Pujolas, and
  Sibiryakov}}]{Blas:2009qj}
\bibinfo{author}{\bibfnamefont{D.}~\bibnamefont{Blas}},
  \bibinfo{author}{\bibfnamefont{O.}~\bibnamefont{Pujolas}}, \bibnamefont{and}
  \bibinfo{author}{\bibfnamefont{S.}~\bibnamefont{Sibiryakov}},
  \bibinfo{journal}{Phys.Rev.Lett.} \textbf{\bibinfo{volume}{104}},
  \bibinfo{pages}{181302} (\bibinfo{year}{2010}), \eprint{0909.3525}.

\bibitem[{\citenamefont{Blas et~al.}(2009)\citenamefont{Blas, Pujolas, and
  Sibiryakov}}]{Blas:2009yd}
\bibinfo{author}{\bibfnamefont{D.}~\bibnamefont{Blas}},
  \bibinfo{author}{\bibfnamefont{O.}~\bibnamefont{Pujolas}}, \bibnamefont{and}
  \bibinfo{author}{\bibfnamefont{S.}~\bibnamefont{Sibiryakov}},
  \bibinfo{journal}{JHEP} \textbf{\bibinfo{volume}{0910}}, \bibinfo{pages}{029}
  (\bibinfo{year}{2009}), \eprint{0906.3046}.

\bibitem[{\citenamefont{Germani et~al.}(2009)\citenamefont{Germani, Kehagias,
  and Sfetsos}}]{Germani:2009yt}
\bibinfo{author}{\bibfnamefont{C.}~\bibnamefont{Germani}},
  \bibinfo{author}{\bibfnamefont{A.}~\bibnamefont{Kehagias}}, \bibnamefont{and}
  \bibinfo{author}{\bibfnamefont{K.}~\bibnamefont{Sfetsos}},
  \bibinfo{journal}{JHEP} \textbf{\bibinfo{volume}{0909}}, \bibinfo{pages}{060}
  (\bibinfo{year}{2009}), \eprint{0906.1201}.

\bibitem[{\citenamefont{Horava and Melby-Thompson}(2011)}]{Horava:2009vy}
\bibinfo{author}{\bibfnamefont{P.}~\bibnamefont{Horava}} \bibnamefont{and}
  \bibinfo{author}{\bibfnamefont{C.~M.} \bibnamefont{Melby-Thompson}},
  \bibinfo{journal}{Gen.Rel.Grav.} \textbf{\bibinfo{volume}{43}},
  \bibinfo{pages}{1391} (\bibinfo{year}{2011}), \eprint{0909.3841}.

\bibitem[{\citenamefont{Janiszewski and Karch}(2012)}]{long}
\bibinfo{author}{\bibfnamefont{S.}~\bibnamefont{Janiszewski}} \bibnamefont{and}
  \bibinfo{author}{\bibfnamefont{A.}~\bibnamefont{Karch}}, \bibinfo{journal}{to
  be published}  (\bibinfo{year}{2012}).

\bibitem[{\citenamefont{Huijse et~al.}(2012)\citenamefont{Huijse, Sachdev, and
  Swingle}}]{Huijse:2011ef}
\bibinfo{author}{\bibfnamefont{L.}~\bibnamefont{Huijse}},
  \bibinfo{author}{\bibfnamefont{S.}~\bibnamefont{Sachdev}}, \bibnamefont{and}
  \bibinfo{author}{\bibfnamefont{B.}~\bibnamefont{Swingle}},
  \bibinfo{journal}{Phys.Rev.} \textbf{\bibinfo{volume}{B85}},
  \bibinfo{pages}{035121} (\bibinfo{year}{2012}), \eprint{1112.0573}.

\bibitem[{\citenamefont{Henkel}(1994)}]{Henkel:1993sg}
\bibinfo{author}{\bibfnamefont{M.}~\bibnamefont{Henkel}},
  \bibinfo{journal}{J.Statist.Phys.} \textbf{\bibinfo{volume}{75}},
  \bibinfo{pages}{1023} (\bibinfo{year}{1994}), \eprint{hep-th/9310081}.

\bibitem[{\citenamefont{Balasubramanian and
  McGreevy}(2008)}]{Balasubramanian:2008dm}
\bibinfo{author}{\bibfnamefont{K.}~\bibnamefont{Balasubramanian}}
  \bibnamefont{and} \bibinfo{author}{\bibfnamefont{J.}~\bibnamefont{McGreevy}},
  \bibinfo{journal}{Phys.Rev.Lett.} \textbf{\bibinfo{volume}{101}},
  \bibinfo{pages}{061601} (\bibinfo{year}{2008}), \eprint{0804.4053}.

\bibitem[{\citenamefont{Goldberger}(2009)}]{Goldberger:2008vg}
\bibinfo{author}{\bibfnamefont{W.~D.} \bibnamefont{Goldberger}},
  \bibinfo{journal}{JHEP} \textbf{\bibinfo{volume}{0903}}, \bibinfo{pages}{069}
  (\bibinfo{year}{2009}), \eprint{0806.2867}.

\bibitem[{\citenamefont{Seiberg}(1997)}]{Seiberg:1997ad}
\bibinfo{author}{\bibfnamefont{N.}~\bibnamefont{Seiberg}},
  \bibinfo{journal}{Phys.Rev.Lett.} \textbf{\bibinfo{volume}{79}},
  \bibinfo{pages}{3577} (\bibinfo{year}{1997}), \eprint{hep-th/9710009}.

\bibitem[{\citenamefont{Sen}(1998)}]{Sen:1997we}
\bibinfo{author}{\bibfnamefont{A.}~\bibnamefont{Sen}},
  \bibinfo{journal}{Adv.Theor.Math.Phys.} \textbf{\bibinfo{volume}{2}},
  \bibinfo{pages}{51} (\bibinfo{year}{1998}), \eprint{hep-th/9709220}.

\bibitem[{\citenamefont{Karch and Katz}(2002)}]{Karch:2002sh}
\bibinfo{author}{\bibfnamefont{A.}~\bibnamefont{Karch}} \bibnamefont{and}
  \bibinfo{author}{\bibfnamefont{E.}~\bibnamefont{Katz}},
  \bibinfo{journal}{JHEP} \textbf{\bibinfo{volume}{0206}}, \bibinfo{pages}{043}
  (\bibinfo{year}{2002}), \eprint{hep-th/0205236}.

\bibitem[{\citenamefont{Karch and O'Bannon}(2007)}]{Karch:2007br}
\bibinfo{author}{\bibfnamefont{A.}~\bibnamefont{Karch}} \bibnamefont{and}
  \bibinfo{author}{\bibfnamefont{A.}~\bibnamefont{O'Bannon}},
  \bibinfo{journal}{JHEP} \textbf{\bibinfo{volume}{0711}}, \bibinfo{pages}{074}
  (\bibinfo{year}{2007}), \eprint{0709.0570}.

\bibitem[{\citenamefont{Ammon et~al.}(2012)\citenamefont{Ammon, Kaminski, and
  Karch}}]{Ammon:2012je}
\bibinfo{author}{\bibfnamefont{M.}~\bibnamefont{Ammon}},
  \bibinfo{author}{\bibfnamefont{M.}~\bibnamefont{Kaminski}}, \bibnamefont{and}
  \bibinfo{author}{\bibfnamefont{A.}~\bibnamefont{Karch}}
  (\bibinfo{year}{2012}), \eprint{1207.1726}.

\bibitem[{\citenamefont{Ogawa et~al.}(2012)\citenamefont{Ogawa, Takayanagi, and
  Ugajin}}]{Ogawa:2011bz}
\bibinfo{author}{\bibfnamefont{N.}~\bibnamefont{Ogawa}},
  \bibinfo{author}{\bibfnamefont{T.}~\bibnamefont{Takayanagi}},
  \bibnamefont{and} \bibinfo{author}{\bibfnamefont{T.}~\bibnamefont{Ugajin}},
  \bibinfo{journal}{JHEP} \textbf{\bibinfo{volume}{1201}}, \bibinfo{pages}{125}
  (\bibinfo{year}{2012}), \eprint{1111.1023}.

\end{thebibliography}

\end{document}